\documentclass[twocolumn,showpacs,preprintnumbers,nofootinbib]{revtex4-1}
\pdfoutput=1

\usepackage{amsfonts}
\usepackage{amsmath}
\usepackage{graphicx}
\usepackage{dcolumn}
\usepackage{bm}
\usepackage{slashed}
\usepackage{amssymb}
\usepackage{MnSymbol}
\usepackage{xcolor}
\usepackage{verbatim}
\usepackage{comment}
\usepackage{todonotes}
\usepackage{grffile}
\usepackage{epstopdf}

\usepackage{soul}

\DeclareMathOperator{\Tr}{Tr}

\begin{document}
\title{The impact of quark masses on pQCD thermodynamics}
\author{Thorben Graf}
\email{tgraf@th.physik.uni-frankfurt.de}
\affiliation{Institute for Theoretical Physics, Goethe University,
Max-von-Laue-Str.\ 1, D--60438 Frankfurt am Main, Germany}
\author{Juergen Schaffner-Bielich}
\email{schaffner@astro.uni-frankfurt.de}
\affiliation{Institute for Theoretical Physics, Goethe University,
Max-von-Laue-Str.\ 1, D--60438 Frankfurt am Main, Germany}
\author{Eduardo S. Fraga}
\email{fraga@if.ufrj.br}
\affiliation{Instituto de Fisica, Universidade Federal do Rio de Janeiro,
Caixa Postal 68528 Rio de Janeiro, RJ 21941-972, Brazil}

%%%%%%%%%%%%%%%%%%%%%%%%%%%%%%%%%%%%%%%%%%%%%%%%%%%%%%%%%%%%%%
\begin{abstract}
We present results for several thermodynamic quantities within the next-to-leading order calculation of the thermodynamic potential in perturbative QCD at finite temperature and chemical potential including non-vanishing quark masses. These results are compared to lattice data and to higher-order optimized perturbative calculations to investigate the trend brought about by mass corrections.
\end{abstract}

\pacs{11.10.Wx,12.38.Bx,12.38.Mh,21.65.Qr}
\keywords{QCD, Phase Diagram of QCD, Quark-Gluon Plasma, NLO Computations}
%%%%%%%%%%%%%%%%%%%%%%%%%%%%%%%%%%%%%%%%%%%%%%%%%%%%%%%%%%%%%%

\maketitle

%%%%%%%%%%%%%%%%%%%%%%%%%%%%%%%%%%%%%%%%%%%%%%%%%%%%%%%%%%%%%%
\section{Introduction}

Ultra-relativistic heavy-ion collision experiments provide enough energy to cross the threshold where hadronic systems are converted into a novel state of matter, the quark-gluon plasma (QGP) \cite{Shuryak:1977ut,Shuryak:1978ij}. In this medium, partons acquire a much longer mean-free path and can move more freely due to asymptotic freedom \cite{Gross:1973id,Gross:1973ju,Gross:1974cs,Politzer:1973fx,Politzer:1974fr}. Below the transition temperature, and up to a few times $T_{c}$, Quantum Chromodynamics (QCD) is essentially nonperturbative. For vanishing baryon chemical potential, Lattice QCD provides an excellent description of the full thermodynamics (see e.g. Refs. \cite{Borsanyi:2015waa,Ding:2015ona} for recent results) that can be compared to improved perturbative calculations for temperatures a few times $T_{c}$ and higher. However, for nonzero baryon densities solid lattice results are still precluded by the well-known Sign Problem \cite{deForcrand:2010ys}, and one has to resort to a combination of perturbative calculations and effective models.

Using perturbative techniques to calculate the thermodynamic potential has a long tradition in QCD \cite{Freedman:1976xs,Freedman:1976dm,Freedman:1976ub,Baluni:1977ms}. Since then, analytic computations of the thermodynamic potential were performed over the years  
(see e.g. \cite{Shuryak:1977ut,Chin:1978gj,Kapusta:1979fh,Toimela:1984xy,Arnold:1994ps,Arnold:1994eb,Zhai:1995ac,Kajantie:2002wa}), and are currently known up to ${\mathcal O}(g^6\ln g)$ in the coupling $g$ at high temperatures and small chemical potentials \cite{Vuorinen:2002ue,Kajantie:2002wa,Vuorinen:2003fs,Ipp:2003jy,Ipp:2003yz,Ipp:2006ij}. It turned out that the pure weak coupling expansion converges badly. The next-to-leading order ($g^2$) shows a surprisingly good agreement with lattice data whereas the next contribution ($g^3$) even causes a flip in the sign of the pressure. It was suggested that the bad convergence is initiated by the contributions of soft momenta, $p_{\text{soft}}\sim gT$. Therefore the pressure of hot QCD is split into two parts $p_{\text{QCD}}\equiv p_{\text{hard}}+p_{\text{soft}}$ \cite{Braaten:1995jr}, where $p_{\text{hard}}\sim \pi T$ are the hard modes. The poor convergence can be assigned to $p_{\text{soft}}$, whereas $p_{\text{hard}}$ exhibits a good convergence. There are ways to reorganize the perturbative series with a special focus on the soft sector. One successful approach is Hard Thermal Loop perturbation theory (HTLpt), see Refs. \cite{Andersen:2009tw,Andersen:2011ug,Andersen:2011sf,Mogliacci:2013mca,Haque:2014rua}. Another method is dimensional reduction \cite{Appelquist:1981vg,Ginsparg:1980ef}, which proceeds by first integrating out the hard modes and leaves an effective three-dimensional theory, dubbed EQCD. A very recent investigation of the Equation of State of quark matter is based on both of the mentioned approaches and guarantees access to all values of temperature and density at order $g^5$ \cite{Kurkela:2016was}. Besides that, at order $g^6$, intrinsic infrared problems appear and can only be solved by non-perturbative methods \cite{Linde:1980ts}.

Although the first calculations of the thermodynamic potential with massive quarks also date back to the 70s \cite{Freedman:1977gz,Kapusta:1979fh,Toimela:1984xy,Farhi:1984qu}, mass effects on the pressure were regarded as being negligible for two decades. The issue was reconsidered at $T=0$ in Ref.~\cite{Fraga:2004gz}, with a modern $\overline{\text{MS}}$ scheme description, and provided corrections that can reach $20$\%, being relevant for the physics of neutron stars. Later, the two-loop result of Ref.~\cite{Fraga:2004gz} was extended to order $g^4$ in Ref.~\cite{Kurkela:2009gj}, which represents the current state-of-the-art perturbative calculation at $T=0$ with massive quarks. The case at zero baryon density and finite temperature was investigated in Ref.~\cite{Laine:2006cp} with consequences to quark mass thresholds (see also Ref.~\cite{Wang:1998tg}). Mass effects on the Yukawa theory were also investigated for massive quarks and mesons in Refs.~\cite{Palhares:2007zzb,Palhares:2008yq,Fraga:2009pi}. In all these cases, results exhibit deviations $\sim 20-30$\%. Quasi-particle models that include finite quark masses were also considered \cite{Peshier:1999ww}.

In this note, we present results for several thermodynamic quantities within the next-to-leading order calculation (NLO) of the thermodynamic potential in perturbative QCD at finite temperature and chemical potential, including non-vanishing quark masses. We are interested in the investigation of effects of consistently introduced massive quarks, which were surprisingly less studied in the past. A model that includes all of these dependences in a reasonable way is the $g^2$-corrected thermodynamic potential with $m_f\neq0$. As already mentioned, it is known for a long time that the pure weak coupling expansion converges badly. However, it is also known that the NLO corrected pressure is relatively close to the lattice data which is of course only accidentally. Nevertheless in terms of a good agreement with lattice simulations the NLO computation provides a baseline from which to study further corrections. What is not known nowadays is the impact of the finite bare quark masses on the thermodynamics. The influence of the strange quark mass for example to the thermodynamical quantities was ignored so far in the literature even though the strange-quark plays an important role, regarding the temperatures achieved in present ultra-relativistic heavy-ion collision experiments. The question is how good is the convergence of the pressure in terms of the quark mass dependence. In Ref.~\cite{Laine:2006cp} it was shown that the convergence is better and the results should be trustable already at NLO. Furthermore, since quark mass effects in $p_{\text{soft}}$ only appear through the thermal masses, Debye-masses $m_{\text{D}}$, bare quark mass effects should be well described in NLO. The only other term that may have a significant quark mass dependence is the NNLO contribution in $p_{\text{hard}}$, but since the convergence of $p_{\text{hard}}$ is better this contribution is small. Indeed, we will show in the following that the model behaves reasonably well in terms of mass and radiative corrections. In order to accomplish a consistent calculation that also includes a running coupling and strange-quark mass, the calculation at $g^2$ is done in the $\overline{\text{MS}}$ renormalization scheme. In principle the calculation of the thermodynamic potential was done by J. I. Kapusta in 1979, see Ref.~\cite{Kapusta:1979fh}. However, it is based on an obsolete renormalization scheme and an explicit numerical evaluation of the thermodynamic potential is missing. \\ We profit from previous calculations of the thermodynamic potential to study other properties such as the entropy, the speed of sound and trace anomaly with massive quarks. These results are compared to lattice data and to higher-order optimized perturbative calculations with the purpose to investigate the trend brought about by mass corrections. Partially included higher order effects that enter our framework by implementing the running of the coupling constant and the strange-quark mass result in an improved description of lattice data already at NLO.

The paper is organized as follows. In section II we summarize the well-established calculation of the thermodynamic potential to this order. In section III we present our results and compare them to lattice data and HTLpt. Section IV contains our summary. Appendix A presents some technical details of the renormalization procedure for nonzero quark masses.

%%%%%%%%%%%%%%%%%%%%%%%%%%%%%%%%%%%%%%%%%%%%%%%%%%%%%%%%%%%%%%
\section{Thermodynamic potential}

The next-to-leading order thermodynamic potential has the diagrammatic form \cite{Kapusta:2006pm}:
\begin{align}
\begin{split}
  \Omega=&-\frac{1}{\beta V}
  \parbox{1.5cm}{
    \includegraphics[width=1.5cm]{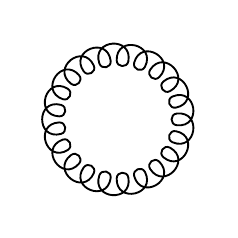}
  }
  +\frac{1}{\beta V}
  \parbox{1.5cm}{
    \includegraphics[width=1.3cm]{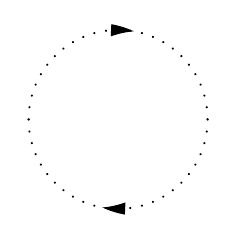}
  }
  +\frac{1}{\beta V}\sum_f
  \parbox{1.5cm}{
    \includegraphics[width=1.3cm]{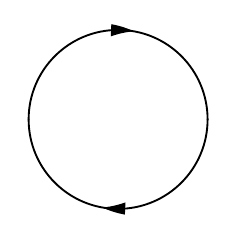}
  }
  \\
  &+\frac{1}{2}\frac{1}{\beta V}\sum_f
  \parbox{1.5cm}{
    \includegraphics[width=1.5cm]{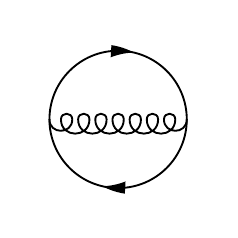}
  }
  +\frac{1}{2}\frac{1}{\beta V}
  \parbox{1.5cm}{
    \includegraphics[width=1.5cm]{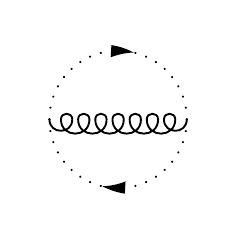}
  }
  \\
  &-\frac{1}{12}\frac{1}{\beta V}
  \parbox{1.35cm}{
    \includegraphics[width=1.35cm]{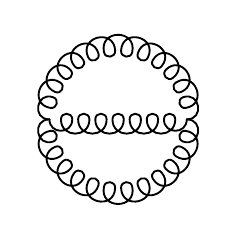}
  }
  -\frac{1}{8}\frac{1}{\beta V}
  \parbox{1.8cm}{
    \includegraphics[width=1.8cm]{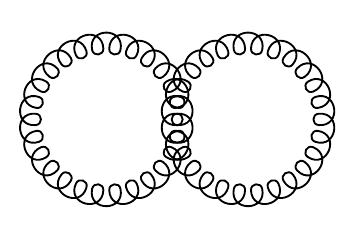}
  }
  \\
  &+\text{ diagrams with counterterms }+\text{ }\mathcal{O}(\text{3 loops}),
  \label{eq-ThermPot}
  \end{split}
\end{align}
where solid lines represent fermions, curly lines gluons and dashed lines ghosts. Notice that including nonzero quark masses, the fermionic contributions can not be rescaled by the flavor number $N_f$. 

The only diagram that will be affected in a nontrivial manner by nonzero quark masses is the exchange diagram. To make the discussion more self contained, we present in some detail the derivation of this contribution. In the Feynman gauge \cite{Kapusta:2006pm}, we have
\begin{align}
\begin{split}
  \label{eq-fermionloop}
  \parbox{2cm}{
    \includegraphics[width=2cm]{Figures/loop1output.pdf}
  }
  =&\frac{1}{2}\beta V N_G g^2\int\frac{d^3p}{(2\pi)^3}\int\frac{d^3q}{(2\pi)^3}\int\frac{d^3k}{(2\pi)^3} \\
   &\times(2\pi)^3\delta(\vec{p}-\vec{q}-\vec{k})T^3\sum_{n_p,n_q,n_k}\beta\delta_{\omega_{n_p},
   \omega_{n_q}+\omega_{n_k}} \\
   &\times\text{Tr}\left[\frac{\gamma_\mu(\slashed{p}+m_f)\gamma^\mu(\slashed{q}+m_f)}{(p^2-m_f^2)k^2(q^2-m_f^2)}\right],
\end{split}
\end{align}
where $N_G=N_c^2-1$ is the number of gluons, $p^\mu=(p^0=i\omega^F_{n_{p}}+\mu,\vec{p})$ is the fermion four-momentum ($q^\mu$ is defined analogously) and $k^\mu=(k^0=i\omega^B_{n_{k}},\vec{k})$ is the gluon four-momentum, with $\omega^B_{n}=2n\pi T$ and $\omega^F_{n}=(2n+1)\pi T$, $n$ being an integer. The mass of the respective quark flavor is $m_f$. The trace is performed in Dirac space. 
It is straightforward to show that Eq.~\eqref{eq-fermionloop} (see e.g. Refs. \cite{Palhares:2008yq,PalharesMaster} for details) reads:
\begin{align}
  \begin{split}
  \label{eq-fermionloop-final}
   &\beta Vg^2\int\frac{d^3pd^3q}{(2\pi)^6}\frac{1}{E_pE_q\omega_{pq}} \\
   &\times\left\{\bar{J}_+\omega_{pq}\Sigma_1+\bar{J}_-\omega_{pq}\Sigma_2\right. \\
   &-\left[\bar{J}_+\left(E_-+\omega_{pq}\right)-\bar{J}_-\left(E_+-\omega_{pq}\right)\right]N_f(p) \\
   &\left.-2\bar{J}_-E_+n_b(\omega_{pq})-\bar{J}_-\left(E_+-\omega_{pq}\right)\right\} \\
   &+\beta Vg^2\int\frac{d^3p}{(2\pi)^3}\frac{N_f(p)}{E_p}\frac{T^2}{6},
  \end{split} 
\end{align}
with the definitions
\begin{equation}
  \begin{split}
      \bar{J}_\pm&\equiv N_G\left[\frac{2m_f^2+\vec{p}\cdot\vec{q}\mp E_pE_q}{E_\mp^2-\omega_{pq}^2}\right], \\
      N_f(p)&\equiv n_f(E_p+\mu)+n_f(E_p-\mu), \\
      N_f(q)&\equiv n_f(E_q+\mu)+n_f(E_q-\mu), \\
      \Sigma_1&\equiv n_f(E_p+\mu)n_f(E_q+\mu)+n_f(E_p-\mu)n_f(E_q-\mu), \\
      \Sigma_2&\equiv n_f(E_p+\mu)n_f(E_q-\mu)+n_f(E_p-\mu)n_f(E_q+\mu),
  \end{split} 
\end{equation}
where $E_\pm\equiv E_p\pm E_q$, $\omega_{pq}\equiv\sqrt{\left|\vec{p}-\vec{q}\right|^2}$, $n_b(\omega)\equiv(\exp(\beta\omega)-1)^{-1}$ and $n_f(E\pm\mu)\equiv(\exp(\beta(E\pm\mu))+1)^{-1}$.

As customary, vacuum contributions can be absorbed by a constant shift of the thermodynamic potential. The ultraviolet-divergent terms 
\begin{align}
  \label{eq-Lf}
    \begin{split}
      L_f=&-\beta Vg^2\int\frac{d^3pd^3q}{(2\pi)^6}\frac{1}{E_pE_q\omega_{pq}} \\
      &\times\left[\bar{J}_+\left(E_-+\omega_{pq}\right)-\bar{J}_-\left(E_+-\omega_{pq}\right)\right]N_f(p),
    \end{split}
\end{align}
\begin{equation}
\label{eq-Lb}
L_b=-\beta Vg^2\int\frac{d^3pd^3q}{(2\pi)^6}\frac{1}{E_pE_q\omega_{pq}}2\bar{J}_-E_+n_b(\omega_{pq}),
\end{equation}
are renormalized in the $\overline{\text{MS}}$ scheme to (see Appendix \ref{sec:renormalization} and Refs.~\cite{Palhares:2008yq,PalharesMaster} for details)
\begin{equation}
L_f^{\textcolor{black}{\text{REN}}}=\frac{\beta V N_Gg^2m_f^2}{4\pi^2}\int \frac{d^3p}{(2\pi)^3}\frac{N_f(p)}{E_p}\left[2+3\ln\left(\frac{\Lambda}{m_f}\right)\right],
\end{equation}
\begin{equation}
L_b^{\textcolor{black}{\text{REN}}}=0,
\end{equation}
so that the renormalized exchange contribution has the form
\begin{align}
  \label{eq-finalexchange}
   \begin{split}
    \Omega_{\rm exch}=&\frac{\alpha_s}{4\pi^3}\int_{m_f}^\infty dE_p\int_{m_f}^\infty dE_q\int_{-1}^1d(\cos\theta) \\
    &\times\sqrt{E_p^2-m_f^2}\sqrt{E_q^2-m_f^2}\{\bar{J}_+\Sigma_1+\bar{J}_-\Sigma_2\} \\
    &+\frac{N_G\alpha_sT^2}{6\pi}\int_{m_f}^\infty dE_p\sqrt{E_p^2-m_f^2}N_f(p) \\
    &+\frac{N_G\alpha_sm_f^2}{4\pi^3}\int_{m_f}^\infty dE_p\sqrt{E_p^2-m_f^2}N_f(p)\left[2+3\ln\left(\frac{\Lambda}{m_f}\right)\right],
   \end{split}
\end{align}
where $\theta$ is the angle between the momenta $\vec{p}$ and $\vec{q}$. Equation~\eqref{eq-finalexchange} recovers previous results \cite{Kapusta:1979fh,Fraga:2004gz} in their limits provided one corrects for the renormalization scheme. Note the explicit scheme dependence at this order due to the nonzero quark mass. The integrals can not be cast in analytic form as in Ref.~\cite{Haque:2014rua} so that they have to be solved numerically.

For the complete thermodynamic potential to this order, one has to add to the exchange contribution the following well-known terms
\begin{equation}
\Omega_{\rm f}=-\frac{N_c}{3\pi^2}\int_m^\infty dE(E^2-m_f^2)^{3/2}N_f(p),
\end{equation}
\begin{equation}
\Omega_{\rm glue}=-\frac{\pi^2}{45}N_GT^4  + \frac{\pi\alpha_s}{36}N_cN_GT^4.
\end{equation}
%

%%%%%%%%%%%%%%%%%%%%%%%%%%%%%%%%%%%%%%%%%%%%%%%%%%%%%%%%%%%%%%
\section{Results}

To investigate the influence of nonzero quark masses for the thermodynamics, we fix the up and down down-quark masses to
\begin{equation}
m_u=2.3 \text{MeV} \quad \text{and} \quad m_d=4.8 \text{MeV},
\label{eq-UDMasses}
\end{equation}
and incorporated the running of the coupling and strange quark mass following Ref.~\cite{Fraga:2004gz}, so that
\begin{equation}
\alpha_s(\Lambda)=\frac{4\pi}{\beta_0L}\left[1-2\frac{\beta_1}{\beta_0^2}\frac{\ln L}{L}\right],
\label{eq-RunAlphas}
\end{equation}
\begin{equation}
m_s(\Lambda)=\hat{m}_s\left(\frac{\alpha_s}{\pi}\right)^{4/9}\left[1+0.895062\frac{\alpha_s}{\pi}\right],
\label{eq-SMass}
\end{equation}
where $L=2\ln(\Lambda/\Lambda_{\overline{\text{MS}}})$, $\beta_0=11-2N_f/3$ and $\beta_1=51-19N_f/3$. The scale $\Lambda_{\overline{\text{MS}}}$ and the invariant mass $\hat{m}_s$ are fixed by requiring $\alpha_s\simeq0.3$ and $m_s\simeq100$ MeV at $\Lambda=2$ GeV \cite{Eidelman:2004wy}; one obtains $\Lambda_{\overline{\text{MS}}}\simeq380$ MeV and $\hat{m}_s\simeq262$ MeV. With these conventions, the only freedom left is the choice of $\Lambda$. The renormalization scale is usually chosen to be $\Lambda=2\sqrt{(\pi T)^2+\mu^2}$. The band uncertainties in our plots shown below result from the variation of $\Lambda$ by a factor of two. In all calculations the quark chemical potentials are set to be equal $\mu_f=\mu=\mu_B/3$. In order to cross check our computations the results are compared to the case of zero temperature and zero chemical potential.

%%%%%%%%%%%%%%%%%%%%%%%%%%%%
\subsection{Pressure, energy density, entropy and trace anomaly}

We start with the zero-temperature case. The one massive flavor case can be computed analytically, yielding, in the $\overline{\text{MS}}$ scheme:
\begin{align}
  \label{eq-ThermPotTZero}
  \begin{split}
    \Omega^{(0)}=&-\frac{N_c}{12\pi^2}\left[\mu u\left(\mu^2-\frac{5}{2}m_f^2\right)+\frac{3}{2}m_f^4\ln\left(\frac{\mu+u}{m_f}\right)\right], \\
    \Omega^{(1)}=&\frac{\alpha_sN_G}{16\pi^3}\left\{3\left[m_f^2\ln\left(\frac{\mu+u}{m_f}\right)-\mu u\right]^2-2u^4\right. \\
    &\left.+m_f^2\left[6\ln\left(\frac{\Lambda}{m_f}\right)+4\right]\left[\mu u-m_f^2\ln\left(\frac{\mu+u}{m_f}\right)\right]\right\},
  \end{split}
\end{align}
where $u\equiv\sqrt{\mu^2-m_f^2}$. The pressure is given by $p=-\Omega$.

\begin{center}
  \begin{figure}
    %\begin{minipage}[c]{8cm}
      \includegraphics[width=8cm]{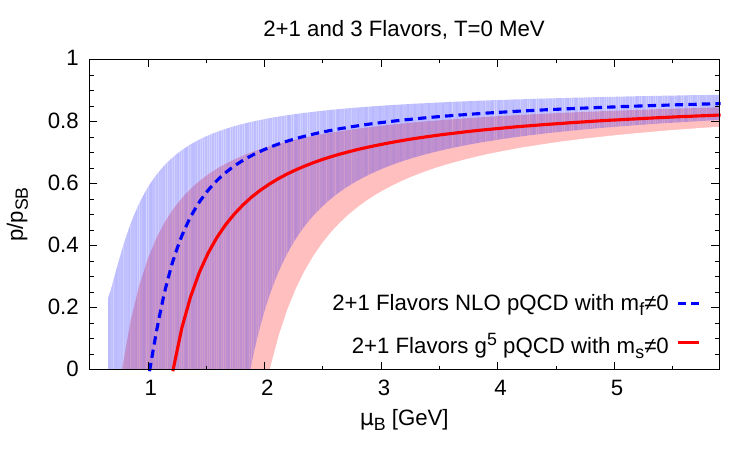}
      \caption{Comparison of the normalized pressure for different orders in perturbation theory; $m_f\neq0$ means that quark masses were chosen according to Eq.~\eqref{eq-UDMasses} and Eq.~\eqref{eq-SMass}. The red band is taken from Ref.~\cite{Kurkela:2009gj} (see also Ref.~\cite{Fraga:2013qra})}
      \label{fig:1}
    %\end{minipage}
  \end{figure}  
\end{center}

Figure~\ref{fig:1} shows a comparison of calculations for the pressure at different orders for 2+1 flavors at vanishing temperature. As customary, we plot the Stefan-Boltzmann-normalized (SB) pressure. The red band is taken from Ref.~\cite{Kurkela:2009gj} (see also Ref.~\cite{Fraga:2013qra}), where the authors assumed massless light quarks but a non-vanishing strange quark mass (The light quark mass effects are, of course, expected to by minor.). This plot quantifies the effect from higher order contributions. As expected, at vanishing temperature they are less severe, even though the pressure is non-ideal even for very high densities.

\begin{center}
  \begin{figure}%[H]
    \begin{minipage}[c]{8cm}
      \includegraphics[width=8cm]{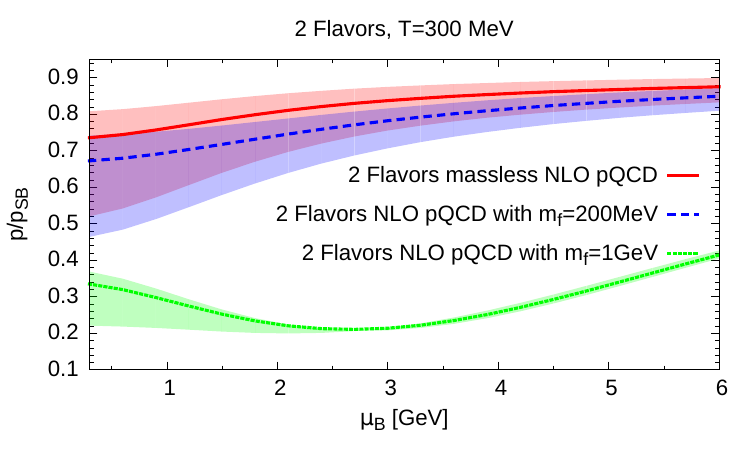}
    \end{minipage}
    \begin{minipage}[c]{8cm}
      \includegraphics[width=8cm]{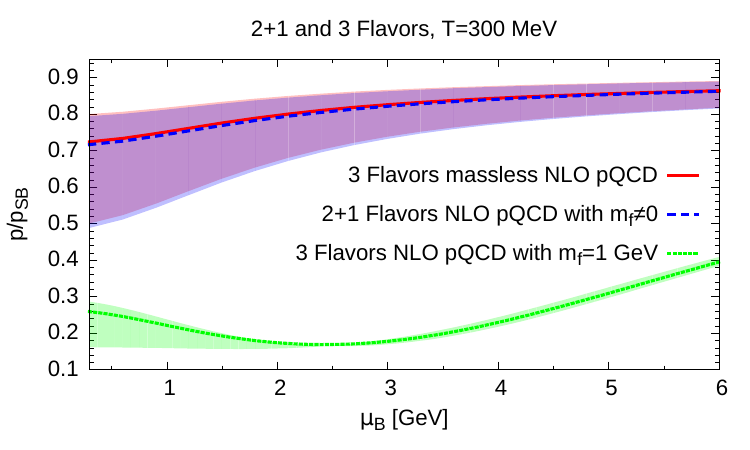}
    \end{minipage}
  \caption{Normalized pressure for 2 (top) and 2+1 / 3 flavors (bottom) for different quark masses; $m_f\neq0$ means that the quark masses were chosen according to Eq.~\eqref{eq-UDMasses} and Eq.~\eqref{eq-SMass}.}
  \label{fig:2}
  \end{figure}  
\end{center}

In Figure~\ref{fig:2} we show the normalized pressure for different numbers of flavors varying the baryon chemical potential and keeping the temperature fixed at $T=300$ MeV. We illustrate the results for different quark masses: the upper plot in Fig.~\ref{fig:2} shows the difference in the normalized pressure $p/p_{SB}$ between the results for massless quarks and those for non-vanishing quark masses with $m_f=200$ MeV (a commonly encountered mass scale in lattice calculations). The difference in the normalized pressure is roughly $\sim 9\%$, so that direct comparisons to lattice data with unphysical quark masses should be done with caution. The lower plot in Fig.~\ref{fig:2} shows the difference in the normalized pressure $p/p_{SB}$ between the massless case and the physical mass case for 2+1 and 3 flavors. This difference in the normalized pressure is less than $1\%$, which shows that temperature effects easily dilute mass effects that are stronger in the cold and dense case (see Refs.~\cite{Fraga:2004gz,Kurkela:2009gj}). Both of the plots in Fig.~\ref{fig:2} unfold the behaviour of the pressure for a number of 2 (top) and 3 (bottom) flavors with all have the same mass of 1 GeV.

\begin{center}
  \begin{figure}%[H]
    \begin{minipage}[c]{8cm}
      \includegraphics[width=8cm]{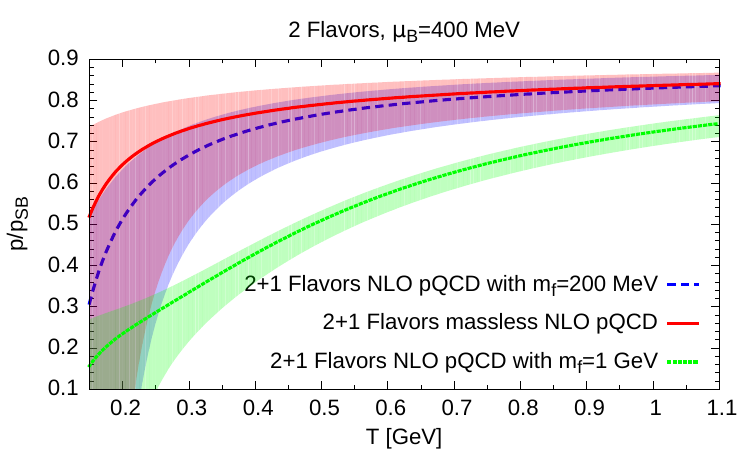}
    \end{minipage}
    \begin{minipage}[c]{8cm}
      \includegraphics[width=8cm]{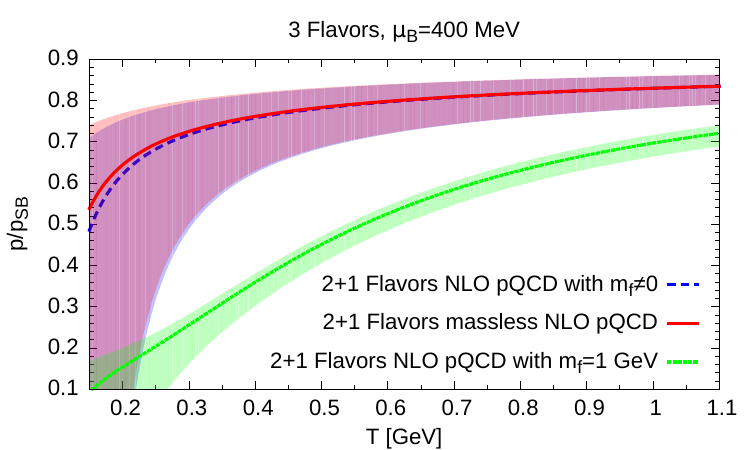}
    \end{minipage}
  \caption{Normalized pressure for 2 (top) and 2+1 / 3 flavors (bottom) for different quark masses; $m_f\neq0$ means that the quark masses were chosen according to Eq.~\eqref{eq-UDMasses} and Eq.~\eqref{eq-SMass}.}
  \label{fig:2a}
  \end{figure}  
\end{center}

We investigate the same situation where the normalized pressure is plotted over the temperature in Fig.~\ref{fig:2a}. This time the difference for the 2-flavor case with increased masses of $m_f=200$ MeV (upper plot in Fig.~\ref{fig:2a}) is larger. For the lowest illustrated temperature of $T=150$ MeV it amounts to almost $\sim 70\%$ at very low temperature. Note that at this temperature and for $\mu_B=400$ MeV $\alpha_s\sim0.54$. For 3 flavors with physical quark masses compared to massless quarks (lower plot in Fig.~\ref{fig:2a}) there is also an increase that is approximately $\sim 11\%$ at $T=150$ MeV and $\mu_B=400$ MeV. In both plots a drastic deviation again appears for increased quark masses of $m_f=1$ GeV.

\begin{center}
  \begin{figure}
    \begin{minipage}[c]{8cm}
      \includegraphics[width=8cm]{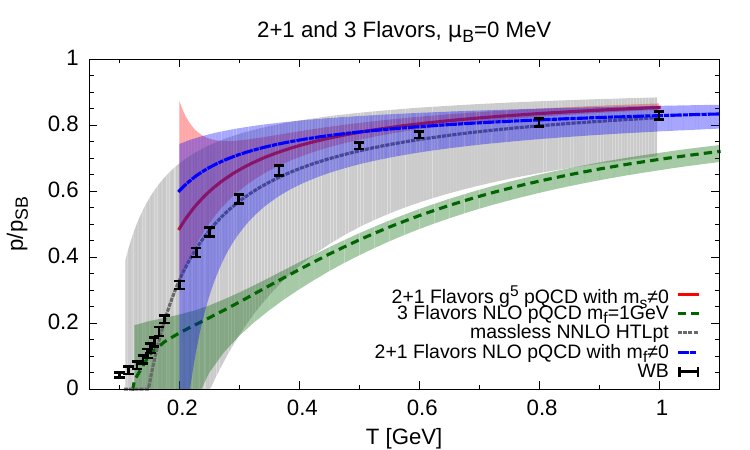}
    \end{minipage}
    \begin{minipage}[c]{8cm}
      \includegraphics[width=8cm]{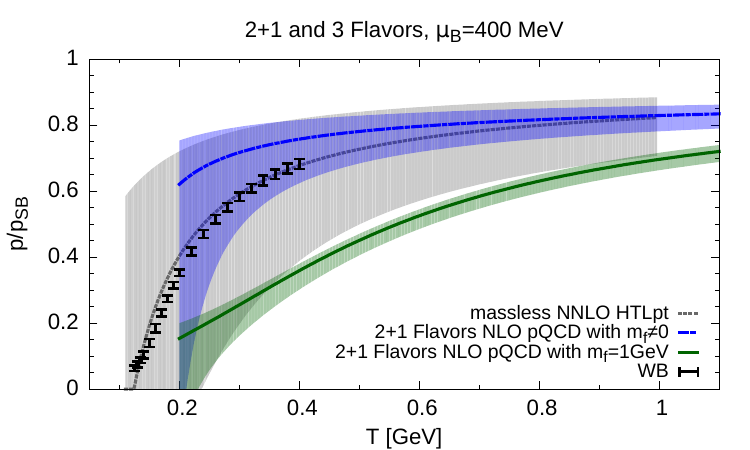}
    \end{minipage}
    \caption{Comparison of the normalized pressure for 2+1 / 3 flavors, $\mu_B=0$ (top) and $\mu_B=400$ MeV (bottom) with HTLpt \cite{Haque:2014rua}, dimensional reduction \cite{Fraga:2013qra}, and lattice results, denoted by WB, \cite{Borsanyi:2010cj,Borsanyi:2012cr}; $m_f\neq0$ means that the quark masses were chosen according to Eq.~\eqref{eq-UDMasses} and Eq.~\eqref{eq-SMass}.}
    \label{fig:3}
  \end{figure}
\end{center}

In Fig.~\ref{fig:3} the normalized pressure for $\mu_B=0$ (top) and for $\mu_B=400$ MeV (bottom) is depicted and compared to lattice simulations \cite{Borsanyi:2012cr} and results of other perturbative investigations that include higher-order effects, for illustration. Of course, higher-order corrections are expected to modify appreciably our simplified description. However, taken as a toy model, this description is not far from lattice results within the interesting regime in temperature. The green bands on the plots of Fig.~\ref{fig:3} again illustrates the large effects that can be brought by the inclusion of heavier quarks.

\begin{center}
  \begin{figure}
    \begin{minipage}[c]{8cm}
      \includegraphics[width=8cm]{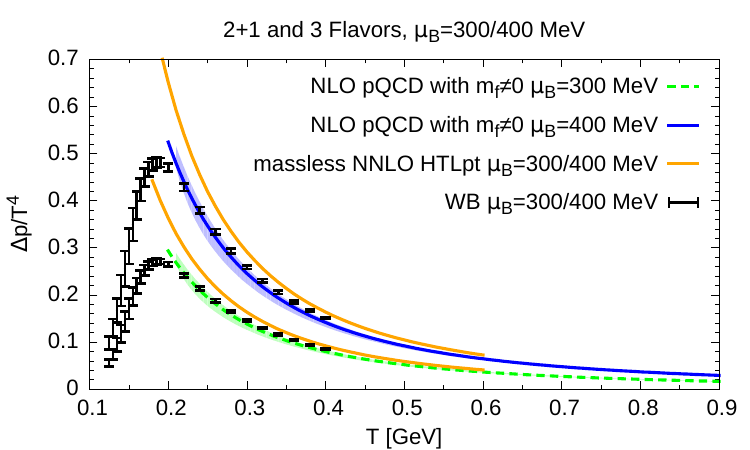}
      \caption{Comparison of the scaled pressure difference with HTLpt \cite{Haque:2014rua} and lattice results, denoted by WB, \cite{Borsanyi:2012cr}; $m_f\neq0$ means that the quark masses were chosen according to Eq.~\eqref{eq-UDMasses} and Eq.~\eqref{eq-SMass}.}
      \label{fig:4}
    \end{minipage}
  \end{figure}
\end{center}

Finally, another comparison can be made by introducing the pressure difference
\begin{equation}
  \Delta p=p(T,\Lambda,\mu)-p(T,\Lambda,0).
  \label{eq-ScaledPressure}
\end{equation}
In Fig.~\ref{fig:4} we show the so-called scaled pressure difference which is the pressure difference, Eq.~\eqref{eq-ScaledPressure}, normalized by $T^4$, which eliminates all contributions to the pressure that do not depend on $\mu$. Notice that lattice data only includes corrections up to order $\mu^2$. Again, taken as a simplified model, our description captures well the dependence on $\mu$ including quark mass effects of this quantity as a function of temperature.

\begin{center}
  \begin{figure}
    \begin{minipage}[c]{8cm}
      \includegraphics[width=8cm]{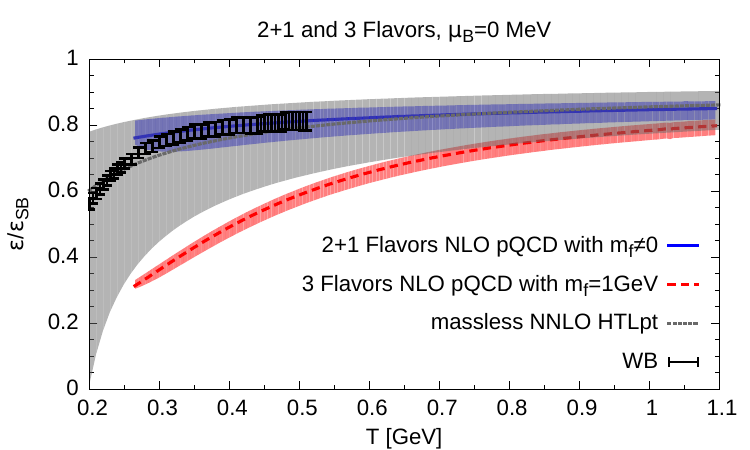}
    \end{minipage}
    \begin{minipage}[c]{8cm}
      \includegraphics[width=8cm]{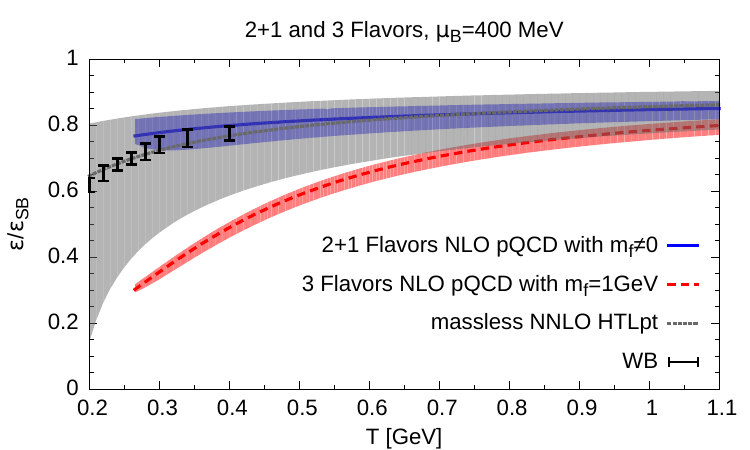}
    \end{minipage}
    \caption{Comparison of the normalized energy density for $\mu_B=0$ (top) and $\mu_B=400$ MeV (bottom) with HTLpt \cite{Haque:2014rua} and lattice results, denoted by WB, \cite{Borsanyi:2010cj,Borsanyi:2012cr}; $m_f\neq0$ means that the quark masses were chosen according to Eq.~\eqref{eq-UDMasses} and Eq.~\eqref{eq-SMass}.}
    \label{fig:5}
  \end{figure}
\end{center}

\begin{center}
  \begin{figure}
    \begin{minipage}[c]{8cm}
      \includegraphics[width=8cm]{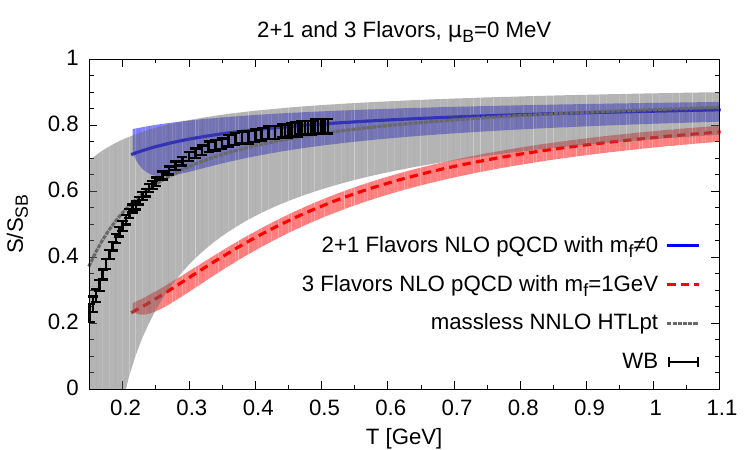}
    \end{minipage}
    \begin{minipage}[c]{8cm}
      \includegraphics[width=8cm]{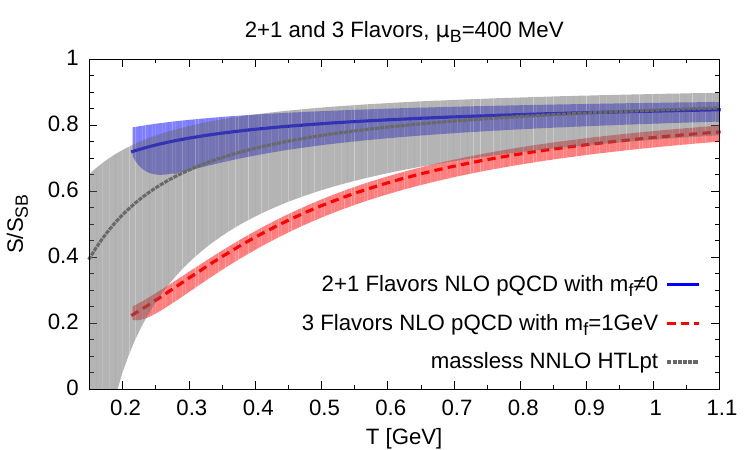}
    \end{minipage}
    \caption{Comparison of the normalized entropy for $\mu_B=0$ (top) and $\mu_B=400$ MeV (bottom) with HTLpt \cite{Haque:2014rua} and lattice results, denoted by WB, \cite{Borsanyi:2010cj,Borsanyi:2012cr}; $m_f\neq0$ means that the quark masses were chosen according to Eq.~\eqref{eq-UDMasses} and Eq.~\eqref{eq-SMass}.}
    \label{fig:6}
  \end{figure}
\end{center}

One can calculate the energy density from the expression of the pressure potential using the relation
\begin{equation}
  \varepsilon=\frac{\partial p}{\partial T}T+\frac{\partial p}{\partial \mu}\mu-p.
  \label{eq-EnergyDensity}
\end{equation}
Note that we fix $\Lambda$ first and then take the derivativs, in contrast to e.g. Ref.~\cite{Haque:2014rua}, where the derivatives are taken first and $\Lambda$ is fixed afterwards. In Fig.~\ref{fig:5} we plot the normalized energy density versus the temperature for vanishing baryon chemical potential (top) and for $\mu_B=400$ MeV (bottom). The interpretation of the results is analogous to the one for the pressure. Here we also show the effect from a heavier quark to the energy density. Finally, the entropy can be directly computed from the pressure, $S=\partial p/\partial T$, and is exhibited in Fig.~\ref{fig:6} for $\mu_B=0$ (top) and $\mu_B=400$ MeV (bottom).

\begin{center}
  \begin{figure}
    \begin{minipage}[c]{8cm}
      \includegraphics[width=8cm]{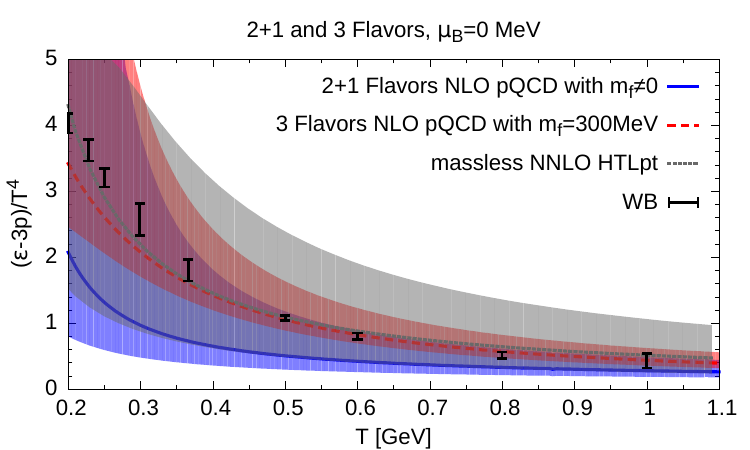}
    \end{minipage}
    \begin{minipage}[c]{8cm}
      \includegraphics[width=8cm]{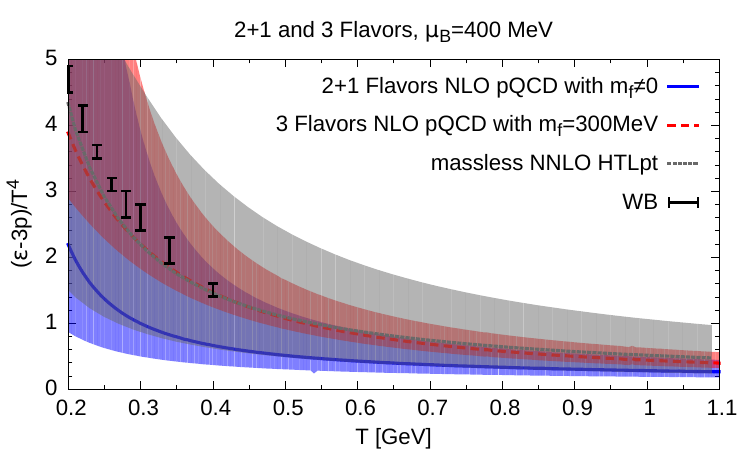}
    \end{minipage}
    \caption{Comparison of the trace anomaly for $\mu_B=0$ (top) and $\mu_B=400$ MeV (bottom) with HTLpt \cite{Haque:2014rua} and lattice results, denoted by WB, \cite{Borsanyi:2010cj,Borsanyi:2012cr}; $m_f\neq0$ means that the quark masses were chosen according to Eq.~\eqref{eq-UDMasses} and Eq.~\eqref{eq-SMass}.}
    \label{fig:7}
  \end{figure}
\end{center}

The trace anomaly of QCD \cite{Andersen:2011ug} is defined as $\mathcal{I}=\varepsilon-3p$. It is related to deconfinement and the gluon condensate \cite{Megias:2009mp}, therefore non-perturbative in its origin. This quantity  vanishes for a massless, noninteracting gas since, in this case, $\varepsilon=3p$. The mass dependence of the results for the trace anomaly could be of special interest since nonzero masses break explicitly the scale invariance of QCD. On the other hand, interactions also break scale invariance as can be seen in the running of $\alpha_s$. We compute $\mathcal{I}$ for 2+1 and 3 flavors, and illustrate it in Fig.~\ref{fig:7} for $\mu_B=0$ (top) and $\mu_B=400$ MeV (bottom). Our results systematically underestimate the trace anomaly, showing that higher order terms are necessary to capture the full running $\alpha_s$ effect. We show also results for a higher quark mass to quantify how it affects this quantity. 

\begin{center}
  \begin{figure}
    \begin{minipage}[c]{8cm}
      \includegraphics[width=8cm]{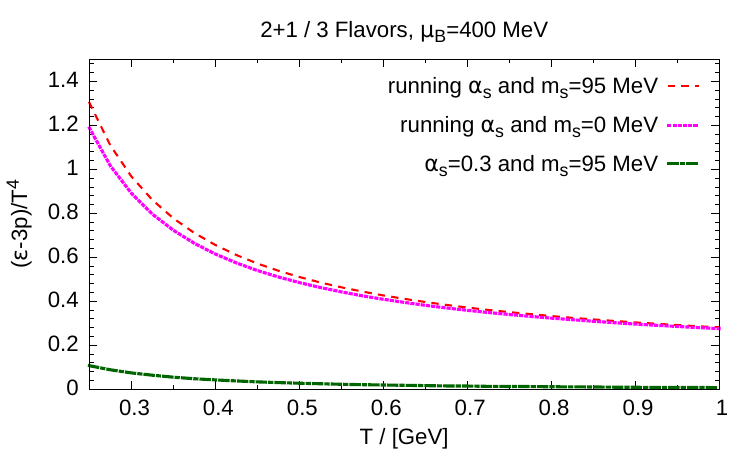}
    \end{minipage}
    \caption{Comparison of the trace anomaly contributions for $\mu_B=400$ MeV}
    \label{fig:7a}
  \end{figure}
\end{center}

The relative contributions of the mass and the running $\alpha_s$ to $\mathcal{I}$ are illustrated in Fig.~\ref{fig:7a}. It is obvious that the one from running $\alpha_s$ clearly dominates. The result for a fixed $\alpha_s$ but a nonvanishing strange quark mass of 95 MeV is rather small. The sum of both contributions is also plotted.

%%%%%%%%%%%%%%%%%%%%%%%%%%%%
\subsection{Speed of sound and susceptibilities}

The speed of sound $c_s$ is given by
\begin{equation}
   \begin{split}
   \label{eq-SoS}
      c_s^2&=\frac{\partial p}{\partial \varepsilon} .%\right|_{S=\text{const.}} .%\\
      %&=\frac{\partial p}{\partial \mu}\frac{1}{\mu\frac{\partial^2p}{\partial\mu^2}}=\frac{\partial p}{\partial T}\frac{1}{T\frac{\partial^2p}{\partial T^2}}.
   \end{split}
\end{equation}  

It is a measure of the stiffness of the equation of state, and is naturally bounded due to causality $v_s\leq1$ and thermodynamic stability $v_s^2>0$. It is relevant at low densities and high temperatures in the hydrodynamical description of the evolution of the quark-gluon plasma formed in relativistic heavy-ion collisions \cite{Jeon:2015dfa,Gale:2013da,Heinz:2013th}. On the other hand, the way the squared speed of sound approaches asymptotically $1/3$ is relevant in the astrophysical context, as discussed in Ref.~\cite{Bedaque:2014sqa}, since there are scenarios within neutron stars which seem to violate $v_s^2<1/3$ (see also Ref.~\cite{Cherman:2009tw}).

\begin{center}
  \begin{figure}
    \begin{minipage}[c]{8cm}
      \includegraphics[width=8cm]{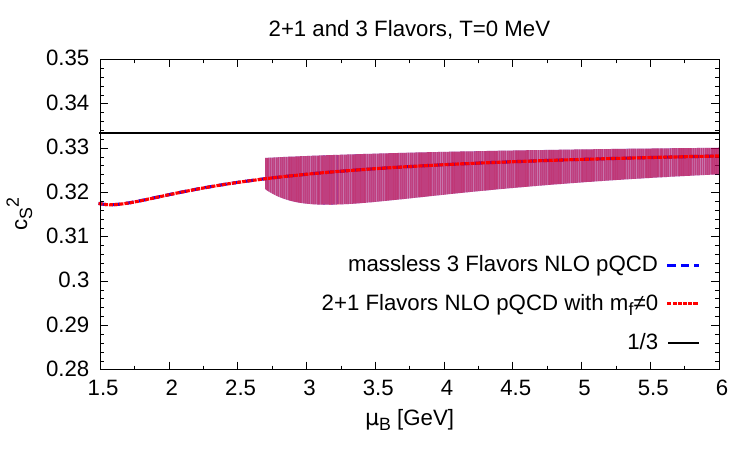}
      \caption{Speed of sound for 2+1 / 3 (massive) flavors at $T=0$; $m_f\neq0$ means that the quark masses were chosen according to Eq.~\eqref{eq-UDMasses} and Eq.~\eqref{eq-SMass}.}
      \label{fig:8}
    \end{minipage}
  \end{figure}  
\end{center}

\begin{center}
  \begin{figure}
    \begin{minipage}[c]{8cm}
      \includegraphics[width=8cm]{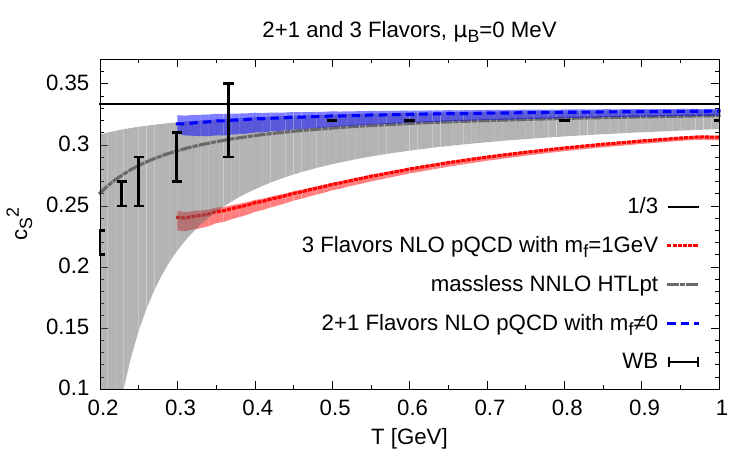}
    \end{minipage}
    \begin{minipage}[c]{8cm}
      \includegraphics[width=8cm]{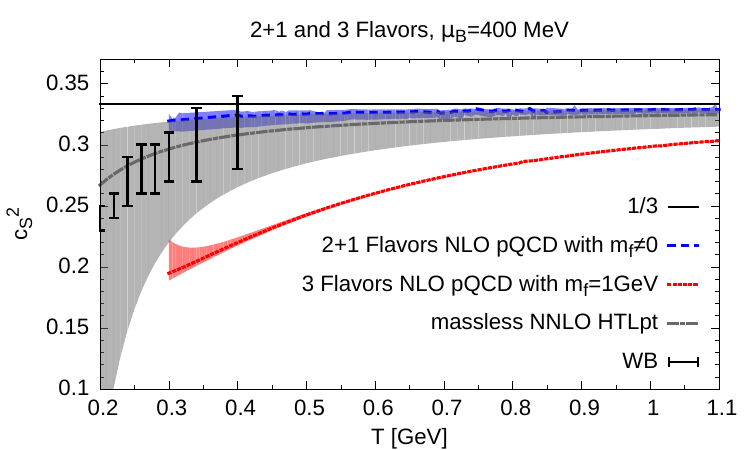}
    \end{minipage}
    \caption{Comparison of the speed of sound for $\mu_B=0$ (top) and $\mu_B=400$ MeV (bottom) with HTLpt \cite{Haque:2014rua}, and lattice results, denoted by WB, \cite{Borsanyi:2010cj,Borsanyi:2012cr}; $m_f\neq0$ means that the quark masses were chosen according to Eq.~\eqref{eq-UDMasses} and Eq.~\eqref{eq-SMass}.}
    \label{fig:9}
  \end{figure}
\end{center}

The speed of sound at zero temperature is shown in Fig.~\ref{fig:8}, which shows that physical quark mass effects and a running coupling constant do not result in a sizable shift from the limiting value $c_s^2=1/3$. In Fig.~\ref{fig:9}, we show the computation of $c_s^2$ for non-vanishing temperature at $\mu_B=0$ (top) and $\mu_B=400$ MeV (bottom). Again, the effect from nonzero masses become relevant only for heavier quarks. The band that is caused by the variation of $\Lambda$ by a factor of two, is only plotted down to densities of $\mu_B\sim2,7$ GeV. Below that density the lower line of the band starts to increase rapidly, which is of course no physical effect, because $\alpha_s$ does so due to the logarithm in $L=2\ln(\Lambda/\Lambda_{\overline{\text{MS}}})$, see Eq.~\eqref{eq-RunAlphas}.
 
We also compute quark number susceptibilities, which are defined as derivatives with respect to the corresponding quark chemical potentials $\vec{\mu}\equiv(\mu_u,\mu_d,\dots,\mu_{N_f})$ as
\begin{equation}
  \left.\chi_{ijk\dots}(T)\equiv\frac{\partial^{i+j+k+\dots}p(T,\vec{\mu})}{\partial\mu^i_u\partial\mu^j_d\partial\mu^k_s\dots}\right|_{\vec{\mu}=0}.
\end{equation}
For instance, $\chi_u^4\equiv\chi_{uuuu}=\frac{\partial^4p}{\partial\mu^4_u}$.

\begin{center}
  \begin{figure}
    \begin{minipage}[c]{8cm}
      \includegraphics[width=8cm]{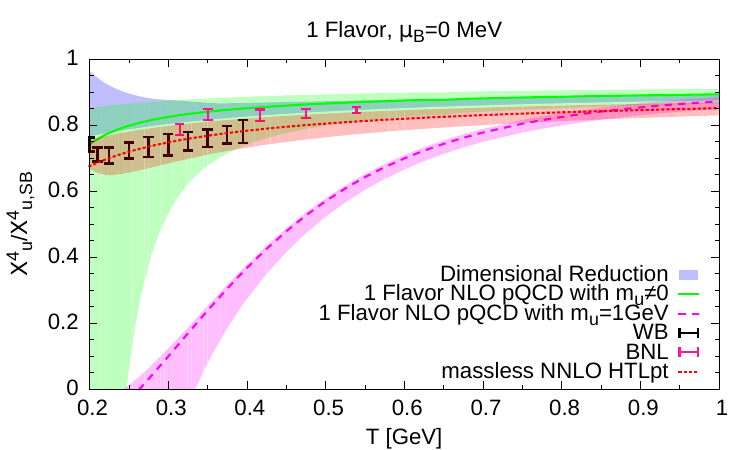}
      \caption{Comparison of the normalized 4th order u-quark susceptibility with HTLpt \cite{Haque:2014rua}, dimensional reduction \cite{Mogliacci:2013mca} and lattice results, denoted by WB, \cite{Borsanyi:2010cj} and BNL, \cite{Bazavov:2013uja}; $m_f\neq0$ means that the quark masses were chosen according to Eq.~\eqref{eq-UDMasses} and Eq.~\eqref{eq-SMass}.}
      \label{fig:10}
    \end{minipage}
  \end{figure}  
\end{center}

\begin{center}
  \begin{figure}
    \begin{minipage}[c]{8cm}
      \includegraphics[width=8cm]{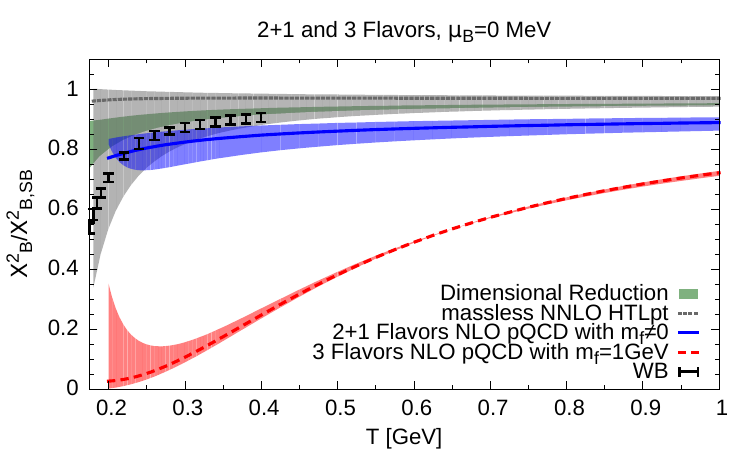}
     \end{minipage}
    \begin{minipage}[c]{8cm}
      \includegraphics[width=8cm]{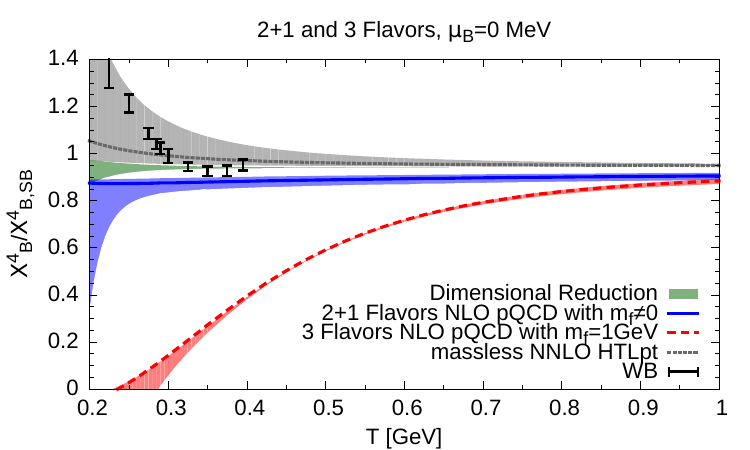}
    \end{minipage}
    \caption{Comparison of the second (top) and fourth (bottom) order baryon number susceptibility with HTLpt, \cite{Haque:2014rua}, dimensional reduction, \cite{Mogliacci:2013mca}, and lattice results, denoted by WB, \cite{Borsanyi:2010cj}; $m_f\neq0$ means that the bare quark masses were chosen according to Eq.~\eqref{eq-UDMasses} and Eq.~\eqref{eq-SMass}.}
    \label{fig:11}
  \end{figure}  
\end{center}

In Fig.~\ref{fig:10} we plot $\chi^4_u$ normalized to its SB-limit and compare it with several other data. Additionally we plot the case for 3 flavors with equal masses of $1$ GeV. In Fig.~\ref{fig:11} the second- (top) and the fourth-order (bottom) baryon number susceptibilities, given by
\begin{equation}
  \left.\chi^n_{B}(T)\equiv\frac{\partial^{n}p}{\partial\mu^n_B}\right|_{\mu_B=0},
\end{equation}
are displayed. Whereas for physical quark masses the mass effect on the susceptibilities is arguably small, for heavier quarks the effect is dramatic, even at comparable and large values of the temperature. Since these quantities seem to be much more sensitive to mass effects, comparisons to lattice results with unphysical high quark masses should be made with caution.

%%%%%%%%%%%%%%%%%%%%%%%%%%%%%%%%%%%%%%%%%%%%%%%%%%%%%%%%%%%%%%
\section{Summary}

We presented a systematic study of the effects from nonzero quark masses on the thermodynamics of perturbative QCD to next-to-leading order at non-vanishing $\mu$ and $T$. We investigated the pressure, scaled pressure difference, energy density, entropy, trace anomaly, speed of sound and susceptibilities and compared our findings to Hard Thermal Loop and Dimensional Reduction calculations at vanishing quark masses and recent lattice data at physical quark masses.

A comparison between perturbative results with vanishing quark masses and lattice results with increased (unphysical) quark masses has a limited validity meaning that the impact of quark masses is not negligible. In this vein, susceptibilities seem to be particularly sensitive to higher quark masses. Although a calculation at higher order (including resummations) is imperative for any quantitative statement, our results point to the role heavier quarks will play when incorporated as in the case of the early universe where charm quarks appear at a few times $T_c$. The interplay between quark masses and isospin is also a relevant issue (see Ref. \cite{Fraga:2008be}) and should be addressed in forthcoming work.

%%%%%%%%%%%%%%%%%%%%%%%%%%%%%%%%%%%%%%%%%%%%%%%%%%%%%%%%%%%%%%
\section*{Acknowledgments}

We would like to thank G. Bergner, N. Haque, R. Stiele, M. Strickland, N. Su and A. Vuorinen for discussions. TG acknowledges specially G. Colucci, L.F. Palhares and P. Petreczky for valuable input. ESF is grateful for the kind hospitality of the ITP group at Frankfurt University, where this work has been initiated. TG is supported by the Helmholtz International Center for FAIR and the Helmholtz Graduate School HGS-HIRe. The work of ESF is partially supported by CAPES, CNPq and FAPERJ.

%%%%%%%%%%%%%%%%%%%%%%%%%%%%%%%%%%%%%%%%%%%%%%%%%%%%%%%%%%%%%%
\begin{appendix}

\section{Renormalization}
\label{sec:renormalization}

We discuss the renormalization of $L_f$ and $L_b$, for details see Ref.~\cite{PalharesMaster}. The expressions derived in \eqref{eq-Lf} and \eqref{eq-Lb} are
\begin{align}
\label{eq-Lf2}
  \begin{split}
    L_f=&-\beta Vg^2\int\frac{d^3pd^3q}{(2\pi)^6}\frac{1}{E_pE_q\omega_{pq}} \\
    &\times\left[\bar{J}_+\left(E_-+\omega_{pq}\right)-\bar{J}_-\left(E_+-\omega_{pq}\right)\right]N_f(p),
  \end{split}
\end{align}
\begin{equation}
\label{eq-Lb2}
  L_b=-\beta Vg^2\int\frac{d^3pd^3q}{(2\pi)^6}\frac{1}{E_pE_q\omega_{pq}}2\bar{J}_-E_+n_b(\omega_{pq}).
\end{equation}
As described in Ref.~\cite{Palhares:2008yq} we define the auxiliary functions
\begin{align}
  \begin{split}
    \mathcal{M}_f(p^4)=&\int_{-\infty}^{+\infty}\frac{dq^4dk^4}{(2\pi)^2}\frac{2m^2-p^\mu q_\mu}{((q^4)^2+E_q^2)((k^4)^2+\omega_{pq}^2)} \\
    &\times2\pi\delta(p^4-q^4-k^4),
  \end{split}
\end{align}
\begin{align}
  \begin{split}
    \mathcal{M}_b(k^4)=&\int_{-\infty}^{+\infty}\frac{dp^4dq^4}{(2\pi)^2}\frac{2m^2-p^\mu q_\mu}{((p^4)^2+E_p^2)((q^4)^2+E_q^2)} \\
    &\times2\pi\delta(p^4-q^4-k^4),
  \end{split}
\end{align}
with $p^\mu=(p^0,\vec{p})=(ip^4,\vec{p})$ and $q^\mu=(q^0,\vec{q})=(iq^4,\vec{q})$. The flavor index of the quark mass $m_f$ is supressed in this appendix. Plugging in the integral representation of the Dirac delta function we factorize and solve the integrations independently \cite{RomatschkeDA}. \\
The function $\mathcal{M}_f(p^4)$ can be written as
\begin{equation}\mathcal{M}_f(p^4)=\frac{1}{4E_q\omega_{pq}}\left[\frac{2m^2+ip^4E_q+\vec{p}\vec{q}}{ip^4+E_q+\omega_{pq}}+\frac{2m^2-ip^4E_q+\vec{p}\vec{q}}{-ip^4+E_q+\omega_{pq}}\right],\end{equation}
and $\mathcal{M}_b(k^4)$ as
\begin{equation}\mathcal{M}_b(k^4)=\frac{2m^2+E_pE_q+\vec{p}\vec{q}}{4E_pE_q}\left[\frac{2E_+}{E_+^2-(ik^4)^2}\right].\end{equation}
Eq.~\eqref{eq-Lf2} and Eq.~\eqref{eq-Lb2} can be expressed in terms of the auxiliary functions as
\begin{align}
  \begin{split}
    L_f&=\beta Vg^2\int\frac{d^3pd^3qd^3k}{(2\pi)^9}(2\pi)^3\delta(\vec{p}-\vec{q}-\vec{k}) \\
    &\times\frac{4N_G}{E_p}N_f(p)\mathcal{M}_f(-iE_p),
  \end{split}
\end{align}
\begin{align}
  \begin{split}
    L_b&=-\beta Vg^2\int\frac{d^3pd^3qd^3k}{(2\pi)^9}(2\pi)^3\delta(\vec{p}-\vec{q}-\vec{k}) \\
    &\times\frac{4N_Gn_b(\omega)}{\omega}\mathcal{M}_b(-i\omega).
  \end{split}
\end{align}
Using the expression for $\mathcal{M}_f(p^4)$ and $\mathcal{M}_b(k^4)$ one arrives at
\begin{equation}
  \begin{split}
    L_f=&4\beta Vg^2N_G\int\frac{d^3p}{(2\pi)^3}\frac{N_f(p)}{E_p}\left\{\int\frac{d^3qd^3k}{(2\pi)^6}\int\frac{dq^4dk^4}{(2\pi)^2}\right. \\
    &\times(2\pi)^3\delta(\vec{p}-\vec{q}-\vec{k})2\pi\delta(p^4-q^4-k^4) \\
    &\left.\left.\times\frac{2m^2-p^\mu q_\mu}{((q^4)^2+E_q^2)((k^4)^2+\omega^2)}\right\}\right|_{p^4=-iE_p},
  \end{split}
\end{equation}
\begin{equation}
  \begin{split}
    L_b=&-4\beta Vg^2N_G\int\frac{d^3k}{(2\pi)^3}\frac{n_b(\omega)}{\omega}\left\{\int\frac{d^3pd^3q}{(2\pi)^6}\int\frac{dp^4dq^4}{(2\pi)^2}\right. \\
    &\times(2\pi)^3\delta(\vec{p}-\vec{q}-\vec{k})2\pi\delta(p^4-q^4-k^4) \\
    &\left.\left.\times\frac{2m^2-p^\mu q_\mu}{((p^4)^2+E_p^2)((q^4)^2+E_q^2)}\right\}\right|_{k^4=-i\omega}. 
  \end{split}
\end{equation}
With the help of the definitions of $p^\mu$ and $k^\mu=(k^0,\vec{k})=(ik^4,\vec{k})$ it follows that
\begin{equation}
  \begin{split}
    L_f=&4\beta Vg^2N_G\int\frac{d^3p}{(2\pi)^3}\frac{N_f(p)}{E_p}\left\{\int\frac{d^4qd^4k}{(2\pi)^8}\right. \\
    &\times(-1)(2\pi)^4i\delta^{(4)}(p^\mu-q^\mu-k^\mu) \\
    &\left.\left.\times\frac{2m^2-p^\mu q_\mu}{((q^4)^2+E_q^2)((k^4)^2+\omega^2)}\right\}\right|_{p^0=E_p},
  \end{split}
\end{equation}
\begin{equation}
  \begin{split}
    L_b=&-4\beta Vg^2N_G\int\frac{d^3k}{(2\pi)^3}\frac{n_b(\omega)}{\omega}\left\{\int\frac{d^4pd^4q}{(2\pi)^8}\right. \\
    &\times(-1)(2\pi)^4i\delta^{(4)}(p^\mu-q^\mu-k^\mu) \\
    &\left.\left.\times\frac{2m^2-p^\mu q_\mu}{((p^4)^2+E_p^2)((q^4)^2+E_q^2)}\right\}\right|_{k^0=\omega}.
  \end{split}
\end{equation}
The expressions in the curly brackets are just the amputated (AMP) self-energy diagrams of a quark or a gluon, respectively, in the vacuum (VAC) where their momenta are put on-shell (M.S.):
\begin{align}
 \begin{split}
    \Bigg( \,
    \parbox{2.5cm}{
    \includegraphics[width=2.5cm]{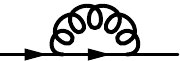}
    }
    \, \Bigg)^{\text{VAC}}_{\substack{\text{AMP} \\ \text{M.S.}}}=&\left\{-g^2\displaystyle\int\frac{d^4qd^4k}{(2\pi)^8}\right. \\
    &\times(2\pi)^4\delta^{(4)}(p^\mu-q^\mu-k^\mu) \\
    &\times\left.\left.\frac{N_G}{N_c}\frac{2m^2-p^\mu q_\mu}{m(q^2-m^2)k^2}\right\}\right|_{\slashed p=m},
 \end{split}
\end{align}
\begin{align}
 \begin{split}
    \Bigg( \,
    \parbox{2.5cm}{
    \includegraphics[width=2.5cm]{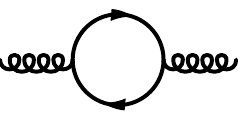}
    }
    \, \Bigg)^{\text{VAC}}_{\substack{\text{AMP} \\ \text{M.S.}}}=&\left\{-g^2\displaystyle\int\frac{d^4pd^4q}{(2\pi)^8}\right. \\
    &\times(2\pi)^4\delta^{(4)}(p^\mu-q^\mu-k^\mu) \\
    &\times\left.\left.N_GN_f\frac{8m^2-4p^\mu q_\mu}{(p^2-m^2)(q^2-m^2)}\right\}\right|_{k^2=0}.
 \end{split}
\end{align}
Defining $p_{\text{new}}^\mu\equiv(i\omega_{n_p}^F+\mu,\vec{p})$ and $q_{\text{new}}^\mu\equiv(i\omega_{n_q}^B,\vec{q})$ with
\begin{center}
  \begin{equation}
    \begin{split}
      T\sum_{n_p}\Tr\left[\frac{1}{\slashed p_{\text{new}}-m}\right]&=4m\frac{N_f(p)-1}{2E_p}, \\
      T\sum_{n_q}\frac{1}{q^2_{\text{new}}}&=-\frac{2n_b+1}{2\omega},
    \end{split}
  \end{equation}
\end{center}
the vacuum self-energies can be written as
\begin{align}
  \begin{split}
     &\displaystyle\sumint_{p_{\text{new}}^\mu}(-1)\Tr\Bigg[\frac{1}{\slashed p_{\text{new}}-m}\Bigg(
     \parbox{2.5cm}{
     \includegraphics[width=2.5cm]{Figures/leticiastyleoutput.pdf}
     }
     \Bigg)^{\text{VAC}}_{\substack{\text{AMP} \\ \text{M.S.}}}\Bigg] \\
     =&g^2\displaystyle\int\frac{d^3p}{(2\pi)^3}\left[\frac{N_f(p)-1}{2E_p}\right] \\
	&\times\left\{\displaystyle\int\frac{d^4qd^4k}{(2\pi)^8}(2\pi)^4\delta^{(4)}(p^\mu-q^\mu-k^\mu)\right. \\
	&\times\left.\left.\frac{N_G}{N_c}\frac{8m^2-4p^\mu q_\mu}{(q^2-m^2)k^2}\right\}\right|_{\slashed p=m},
  \end{split}
\end{align}
\begin{align}
  \begin{split}
     &\displaystyle\sumint_{q_{\text{new}}^\mu}\frac{g_{\mu\nu}}{q^2_{\text{new}}}\Bigg(
     \parbox{2.5cm}{
     \includegraphics[width=2.5cm]{Figures/gluonwithfermionloopoutput.pdf}
     }
     \Bigg)^{\text{VAC}}_{\substack{\text{AMP} \\ \text{M.S.}}} \\
     =&g^2\displaystyle\int\frac{d^3k}{(2\pi)^3}\left[\frac{2n_b+1}{2\omega}\right] \\
	&\times\left\{\displaystyle\int\frac{d^4pd^4q}{(2\pi)^8}(2\pi)^4\delta^{(4)}(p^\mu-q^\mu-k^\mu)\right. \\
	&\times\left.\left.N_GN_f\frac{8m^2-4p^\mu q_\mu}{(p^2-m^2)(q^2-m^2)}\right\}\right|_{k^2=0}.
  \end{split}
\end{align}
Additionally, one can transform it into the form:
\begin{align}
  \begin{split}
     L_f=&-2\beta VN_ci\textcolor{black}{\Bigg\{}\displaystyle\sumint_{p_{\text{new}}^\mu}(-1) \\
     &\times\Tr\Bigg[\frac{1}{\slashed p_{\text{new}}-m}\Bigg(
     \parbox{2.5cm}{
     \includegraphics[width=2.5cm]{Figures/leticiastyleoutput.pdf}
     }
     \Bigg)^{\text{VAC}}_{\substack{\text{AMP} \\ \text{M.S.}}}\Bigg]\textcolor{black}{\Bigg\}_{\text{MAT}}},
  \end{split}
\end{align}
\begin{align}
  \begin{split}
     L_b=\displaystyle\frac{\beta V}{N_f}i\textcolor{black}{\Bigg\{}\sumint_{q_{\text{new}}^\mu}\frac{g_{\mu\nu}}{q_{\text{new}}^2}\Bigg(
     \parbox{2.5cm}{
     \includegraphics[width=2.5cm]{Figures/gluonwithfermionloopoutput.pdf}
     }
     \Bigg)^{\text{VAC}}_{\substack{\text{AMP} \\ \text{M.S.}}}\textcolor{black}{\Bigg\}_{\text{MAT}}},
  \end{split}
  \label{eq-ClosedBVE}
\end{align}
whereupon the MAT index denotes that we take only the matter part of the expressions in the curly brackets. This means that the pure vacuum part has already been subtracted and, while $p^\mu$ and $k^\mu$ are evaluated on the mass shell, $p_{\text{new}}^\mu$ and $k_{\text{new}}^\mu$ are not. These expressions can be assigned to the following two two-loop-diagrams
\begin{align}
  \begin{split}
     L_f\quad\longmapsto\quad
     \parbox{2.5cm}{
     \includegraphics[width=2.5cm]{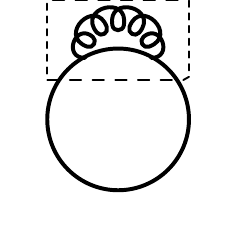}
     },
  \end{split}
\end{align}

\begin{align}
  \begin{split}
     L_b\quad\longmapsto\quad
     \parbox{2.5cm}{
     \includegraphics[width=2.5cm]{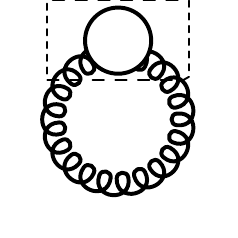}
     }.
  \end{split}
\end{align}
The dashed box within the right diagrams indicates that this part of the diagram is put on-shell. Hence the UV-divergent part is completely covered by the dashed box. The matter part of the remaining integral does not contain any new divergences. \\
Therefore the renormalization procedure reduces to the renormalization of the one-loop diagram in the vacuum. In the $\overline{\text{MS}}$ scheme the final renormalized expression for the amputated quark self-energy is:
\begin{align}
  \begin{split}
    &\Bigg(
    \parbox{2.5cm}{
    \includegraphics[width=2.5cm]{Figures/leticiastyleoutput.pdf}
    }
    \Bigg)^{\text{VAC}}_{\substack{\text{AMP} \\ \text{M.S.,} \textcolor{black}{\text{REN}}}}\displaystyle \\
    =&-\frac{img^2N_G}{(4\pi)^2N_c}\int_0^1dx\left\{\ln\left(\frac{\Lambda^2}{\Delta}\right)(2-x)+(x-1)\right\},
  \end{split}
\end{align}
where $\Delta=m^2(1-x)^2$ and $\Lambda$ is the renormalization scale. Finally one finds that
\begin{equation}L_f^{\textcolor{black}{\text{REN}}}=\frac{\beta V N_Gg^2m^2}{4\pi^2}\int \frac{d^3p}{(2\pi)^3}\frac{N_f(p)}{E_p}\left[2+3\ln\left(\frac{\Lambda}{m}\right)\right].\end{equation}
The renormalization of $L_b$ poceeds analogeously but turns out to vanish. The general Lorentz structure of a bosonic self-energy in the vacuum reads
\begin{equation}
\label{eq-BosonicLS}
  \Pi_{\mu\nu}(k)=\left(k^2g_{\mu\nu}-k_\mu k_\nu\right)\Pi(k^2).
\end{equation}
In Eq.~\eqref{eq-ClosedBVE} the bosonic self-energy is multiplied by $g_{\mu\nu}$ which contracts with the Lorentz structure in Eq.~\eqref{eq-BosonicLS}. The momentum of the ingoing gluon must be put on-shell, hence for a massless gluon $k^2=0$. Subsequently, the whole expression vanishes because it is proportional to $k^2$.

\end{appendix}

\bibliographystyle{ieeetr}
\bibliography{pQCD_MassiveQuarks_EPJA_Refs}

\end{document}